\documentclass[twocolumn]{aastex631}
\usepackage{booktabs}
\usepackage{graphicx}
\usepackage{float}
\usepackage{ulem}
\usepackage[multiple]{footmisc}

\begin{document}

\title{Contribution of Unresolved Sources to Diffuse Gamma-Ray Emission from the Galactic Plane}

\author{Jiayin He}
\affiliation{Key Laboratory of Dark Matter and Space Astronomy, Purple Mountain Observatory, Chinese Academy of Sciences, Nanjing 210023, China}
\affiliation{School of Astronomy and Space Science, University of Science and Technology of China, Hefei 230026, China}

\author{Houdun Zeng}
\affiliation{Key Laboratory of Dark Matter and Space Astronomy, Purple Mountain Observatory, Chinese Academy of Sciences, Nanjing 210023, China}
\affiliation{School of Astronomy and Space Science, University of Science and Technology of China, Hefei 230026, China}
\affiliation{Key Laboratory of Astroparticle Physics of Yunnan Province, Yunnan University, Kunming 650091, China}
\email{zhd@pmo.ac.cn}

\author{Yi Zhang}
\affiliation{Key Laboratory of Dark Matter and Space Astronomy, Purple Mountain Observatory, Chinese Academy of Sciences, Nanjing 210023, China}
\affiliation{School of Astronomy and Space Science, University of Science and Technology of China, Hefei 230026, China}
\email{zhangyi@pmo.ac.cn}

\author{Qiang Yuan}
\affiliation{Key Laboratory of Dark Matter and Space Astronomy, Purple Mountain Observatory, Chinese Academy of Sciences, Nanjing 210023, China}
\affiliation{School of Astronomy and Space Science, University of Science and Technology of China, Hefei 230026, China}

\author{Rui Zhang}
\affiliation{Key Laboratory of Dark Matter and Space Astronomy, Purple Mountain Observatory, Chinese Academy of Sciences, Nanjing 210023, China}
\affiliation{School of Astronomy and Space Science, University of Science and Technology of China, Hefei 230026, China}

\author{Jun Li}
\affiliation{Key Laboratory of Dark Matter and Space Astronomy, Purple Mountain Observatory, Chinese Academy of Sciences, Nanjing 210023, China}
\affiliation{School of Astronomy and Space Science, University of Science and Technology of China, Hefei 230026, China}



\begin{abstract}

The diffuse gamma-ray emission from the Milky Way serves as a crucial probe for understanding the propagation and interactions of cosmic rays within our galaxy. The Galactic diffuse gamma-ray emission between 10 TeV and 1 PeV has been recently measured by the square kilometer array (KM2A) of the Large High Altitude Air Shower Observatory (LHAASO). The flux is higher than predicted for cosmic rays interacting with the interstellar medium. In this work, we utilize a non-parametric method to derive the source count distribution using the published first LHAASO source catalog. Based on this distribution, we calculate the contribution of unresolved sources to the diffuse emission measured by KM2A. When comparing our results to the measured diffuse gamma-ray emission, we demonstrate that for the outer Galactic region, the contributions from unresolved sources and those predicted by models are roughly consistent with experimental observations within the uncertainty. However, for the inner Galactic region, additional components are required to account for the observed data.

\end{abstract}



\section{Introduction} \label{sec:intro}
 The measurement of diffuse Galactic emission (DGE) from the galactic plane is a critical aspect of astrophysics, providing insights into the origin and propagation of cosmic rays (CRs), as well as serving as a probe for the nature of the interstellar medium. Diffuse Galactic emissions can be produced by various mechanisms, including neutral pion decays in the interstellar medium, inverse Compton scattering on radiation fields, and bremsstrahlung \citep{1996On, Strong_2000, Strong_2004}. The diffuse gamma emission was first observed by SAS-2 \citep{1975ApJ...198..163F}, and subsequently investigated by COS-B \citep{1982A&A...105..164M} and EGRET \citep{Hunter_1997}. A detailed survey and measurement of the DGE in the GeV energy band was later provided by Fermi-LAT \citep{Ackermann_2012}. In the TeV energy range and above, space-based detectors face challenges in accumulating a sufficient number of signals.
However, advances in detection technologies in ground-based experiments have enabled the successful detection of the DGE by Milagro \citep{Abdo_2007, Abdo_2008}, HESS \citep{2014PhRvD..90l2007A}, ARGO-YBJ \citep{Bartoli_2015}, Tibet AS$\gamma$ \citep{2021PhRvL.126n1101A}, LHAASO \citep{PhysRevLett.131.151001}, and HAWC \citep{Alfaro_2024}.

 Recent observations of DGE at very high energies (VHE) by ground-based air shower experiments reveal a discrepancy between the observed data and the standard CR propagation models. LHAASO-KM2A reported with high precision the energy spectra of DGE in the range from 10 TeV to 1 PeV \citep{PhysRevLett.131.151001}. The measured flux in the inner Galaxy region ($15^{\circ}<l<125^{\circ}, |b|<5^{\circ}$) is approximately three times higher than model predictions, particularly for the flux between 10 TeV and 60 TeV. The outer Galaxy region ($125^{\circ}<l<235^{\circ}, |b|<5^{\circ}$) also shows a similar result. In addition to the spectral energy distribution (SED) of the DGE, the longitude distribution of the DGE measured by LHAASO-KM2A presents a clear deviation from the gas spatial distribution. Most recently, HAWC performed an analysis of diffuse emission at TeV energies from a region of the Galactic plane, specifically over the longitude range $40^{\circ}<l<70^{\circ}$. The DGE flux observed is higher by about a factor of two compared with the DRAGON-based model \citep{Alfaro_2024}. From 1 to 500 GeV, Fermi-LAT data also shows an excess of DGE, using the same regions of interest (ROIs) as LHAASO-KM2A and subtracting the resolved sources and the isotropic diffuse gamma-ray background \citep{Zhang_2023}.

One interpretation for the origin of the flux excess is that it originates from unresolved sources which are too faint to be detected by current detectors \citep{PhysRevLett.131.151001,Vecchiotti:2021yzk}. A power-law spectrum with an exponential cutoff was found to be able to describe the observed  excess \citep{Zhang_2023}, while another research has proposed the inclusion of two additional components to provide a more comprehensive explanation for the excess \citep{PhysRevD.108.L061305}. TeV halos surrounding pulsars could be a natural explanation of this unresolved source population
\citep{PhysRevLett.120.121101,Yan_2024,dekker2023diffuse}. Alternative explanations include the possibility of confinement and interactions of cosmic rays around acceleration sites \citep{Zhang:2021xri,He_2024,sun2023multimessenger}, the spatially dependent propagation model \citep{Lipari:2018gzn,Luque:2022buq,PhysRevD.109.063001}, the anisotropic diffusion model \citep{Giacinti:2023ljr}, or the signals leakage from the extended known sources \citep{chen2024newperspectivediffusegammaray}.

It is inevitable that $\gamma$-ray sources with fluxes below the detection limit contribute cumulatively to the measured diffuse emission. In this work, we try to estimate the contribution from the unresolved source population, based on the LHAASO observations of bright sources. Using the first LHAASO source catalog \citep{Cao_2024}, we derive the intrinsic distributions of the integral flux and photon spectral index (hereafter referred to as photon index) through a non-parametric method \citep{LedynBell_1971}. From these distributions, we calculate the detection efficiency, which is then used to estimate the flux of unresolved sources.

The structure of this paper is as follows. In Section 2, we describe the data set from the first LHAASO catalog. In Section 3, we derive the distributions of the gamma-ray sources. In Section 4, we estimate the flux from unresolved sources. Section 5 presents the results and compares them with the measurements of LHAASO-KM2A. Finally, Section 6 provides the conclusion and discussion.

\section{Data Set} \label{sec:data}
In this paper, our analysis is based on the first LHAASO catalog \citep{Cao_2024}, which provides information about the spectral distribution and extensions of gamma-ray sources. The LHAASO catalog contains 90 sources with an angular size smaller than $2^{\circ}$, detected by the Water Cherenkov Detector Array (WCDA) and KM2A. Notably, not every source has been detected by both arrays simultaneously. KM2A has detected 75 sources above 25 TeV with a test statistic (TS) greater than 33, while WCDA has observed 69 sources in the range from 1 TeV to 25 TeV with a TS of at least 36. Here, $TS = 2 \times \ln(L_{1}/L_{0})$, where $L_{1}$ and $L_{0}$ represent the likelihoods for the alternative hypothesis (background plus signal) and the null hypothesis (background only), respectively. In the LHAASO catalog analysis, the spectra of sources are assumed to follow separate power-law models for those detected in the WCDA energy range (1-25 TeV) and the KM2A energy range ($>25$  TeV), respectively. The sources considered in this analysis are only those with $|b|<5^{\circ}$, resulting in 65 sources for KM2A and 60 sources for WCDA.

We focus on two crucial parameters of a source: the integrated flux and the photon spectral index. The integrated flux is defined as follows:
\begin{equation}\label{eq:intF}
    F=\int_{E_{\rm min}}^{E_{\rm max}} \phi_{0}\left(\frac{E}{E_{0}}\right)^{-\Gamma}dE,
\end{equation}
where  $\phi_{0}$ represents the differential flux at the pivot energy $E_{0}$, which is 50 TeV for KM2A and 3 TeV for WCDA. $\Gamma$ is the index of the assumed power-law spectrum. These parameters can be obtained from the LHAASO catalog paper. Specifically, for KM2A, $E_{\rm min}=25$ TeV and $E_{\rm max}=1600$ TeV, while for WCDA, $E_{\rm min}=1$ TeV and $E_{\rm max}=25$ TeV. Noted that even if a source is detected by both arrays simultaneously, the value of $\Gamma$ will differ between the two energy ranges.

\begin{figure*}[ht!]
\includegraphics[width=0.96\textwidth]{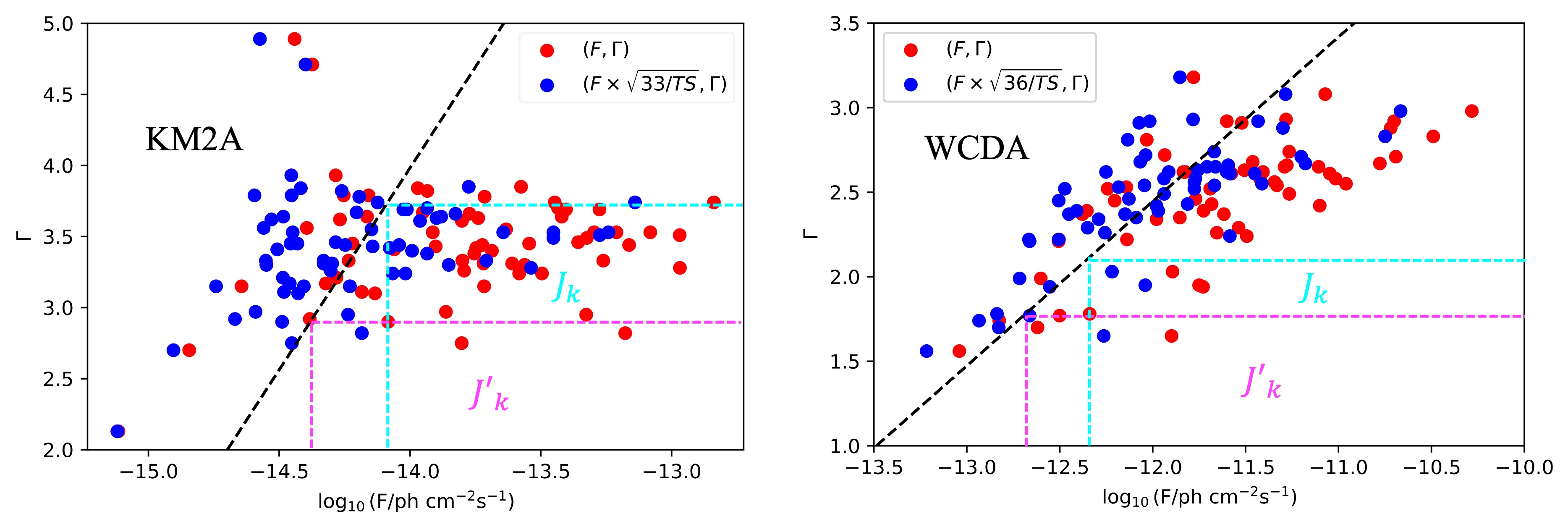}
\caption{Integrated flux and photon spectral index of LHAASO sources with $|b| < 5^{\circ}$ used in this analysis. Left panel: KM2A; right panel: WCDA. There are 65 sources in the KM2A sample and 60 sources in the WCDA sample, represented by the red points with parameters $(F, \Gamma)$. The integrated flux is the integration of the power-law SED with respect to energy, ranging from 25 to 1600 TeV for KM2A and from 1 to 25 TeV for WCDA. The dashed black lines indicate the estimated detection threshold for the first LHAASO source survey, calculated by fitting the blue data points scaled from the red data points based on their TS. The cyan and magenta squares illustrate the data-defined sets $J_{k}$ and $J_{k}'$, respectively.}
\label{fig:gammalgF}
\end{figure*}
The integrated flux and photon index of sources detected by KM2A and WCDA are illustrated in Figure \ref{fig:gammalgF} as red points. This figure underscores the threshold effects resulting from the detection capabilities of both KM2A and WCDA, indicating that sources with harder spectra generally exhibit lower flux levels compared to those with softer spectra \citep{Abdo_2010}. The threshold limit is estimated following the methodology outlined by \citep{Singal_2012}. Initially, we normalize the integrated flux using the formulas $F' = F \times \sqrt{33/TS}$ for KM2A and $F' = F \times \sqrt{36/TS}$ for WCDA, thereby standardizing the fluxes to a uniform TS level. The blue points in Figure \ref{fig:gammalgF} depict these normalized integrated fluxes. Assuming a linear relationship between the integrated flux $F'$ and the photon index $\Gamma$, we apply a linear regression to the ($\Gamma, F'$) points to derive a line representing the detection threshold limit corresponding to the flux value at a minimum TS for a given photon index. These limit lines are depicted as dashed black lines in Figure \ref{fig:gammalgF} for both KM2A and WCDA.

\section{Intrinsic Distribution of Sources} \label{sec:distri}
In this section, we derive the intrinsic distribution of integrated flux and photon index of sources from a truncated sample using a non-parametric method, specifically the Lynden-Bell \( C^{-} \) method \citep{LedynBell_1971}. This method has previously been employed to calculate the intrinsic distributions of luminosity or/and flux and photon index for extragalactic sources \citep[e.g.][]{Singal_2012, Zeng_2021,2024Univ...10..340Y}. The integrated flux and the spectral index are not expected to be intrinsically correlated, as the flux is a distance-dependent measure while the photon index is not. However, the selection process introduces an artificial correlation, as seen in the case of, e.g., WCDA (Figure \ref{fig:gammalgF}). To address this, we apply the Lynden-Bell \( C^{-} \) method to remove the spurious correlation introduced by the selection process.

\subsection{Integrated flux distribution} \label{subsec:INTFLUX}

Firstly, we define a data set \( J_{k} \) for a source labeled as \( k \) with parameters \( (\Gamma_{k}, F_{k}) \) as follows:
\begin{equation}\label{eq:fluxeq}
    J_{k}=\{j|F_{j}\geq F_{k}, \Gamma _{j}\leq \Gamma _{k} ^{max}\}
\end{equation}
Here, \( F_{j} \) represents the integrated flux of the \( j \)-th source, \( \Gamma_{j} \) denotes its photon index, and \( \Gamma_{k}^{\text{max}} \) is the maximum photon index determined by the detection limit line. This line indicates the threshold beyond which a source with integrated flux \( F_{k} \) can be observed. For example, the \( k \)-th source is represented by the intersection of cyan and magenta lines Figure \ref{fig:gammalgF}. Its associated data set \( J_{k} \) is shown as the cyan square in the left panel for the KM2A sample and in the right panel for the WCDA sample. The number of sources within this square is denoted as \( n_{k} \).

The cumulative distribution for integrated flux follows the equation of the non-parametric method \citep{LedynBell_1971}:
\begin{equation}\label{eq:psiL}
    \psi(F_k) = \prod_{j} \left( 1 + \frac{1}{n_{j}-1} \right),
\end{equation}
where this product is computed over all sources with integrated flux \( F_{j} > F_{k} \), and \( n_{j}-1 \) represents the number of sources in data set \( J_{j} \) excluding the \( j \)-th source itself. The cumulative distribution indicates the number of sources with an integrated flux greater than a specific value, $F$, denoted as $N(>F)$, which is defined by:
\begin{equation}\label{eq:intNF}
    N(>F) \equiv \psi(F) = \int_{F}^{\infty} \frac{dN}{dF'} \, dF',
\end{equation}
where \( \frac{dN}{dF'} \) is the differential distribution of integrated flux.

The differential distribution \( \frac{dN}{dF} \) is modeled as a broken power-law:
\begin{equation}\label{eq:flux_diff}
    \frac{dN}{dF} = \left\{ \begin{array}{ll}
    A \left( \frac{F}{F_{0}} \right)^{-\beta_{1}} & \text{for } F < F_{b} \\
    A \left( \frac{F_{b}}{F_{0}} \right)^{-\beta_{1} + \beta_{2}} \left( \frac{F}{F_{0}} \right)^{-\beta_{2}} & \text{for } F \geq F_{b}
    \end{array}\right.
\end{equation}
Here, \( A \) represents the normalization factor at \( F_{0} \), where \( F_{0} = 10^{-14} \ \text{ph} \ \text{cm}^{-2} \ \text{s}^{-1} \) for KM2A and \( F_{0} = 10^{-12} \ \text{ph} \ \text{cm}^{-2} \ \text{s}^{-1} \) for WCDA. \( F_{b} \) denotes the break integrated flux, and \( \beta_{1} \) and \( \beta_{2} \) are the spectral indices below and above \( F_{b} \), respectively.

\begin{figure*}[ht!]
\centering
\includegraphics[width=0.96\textwidth]{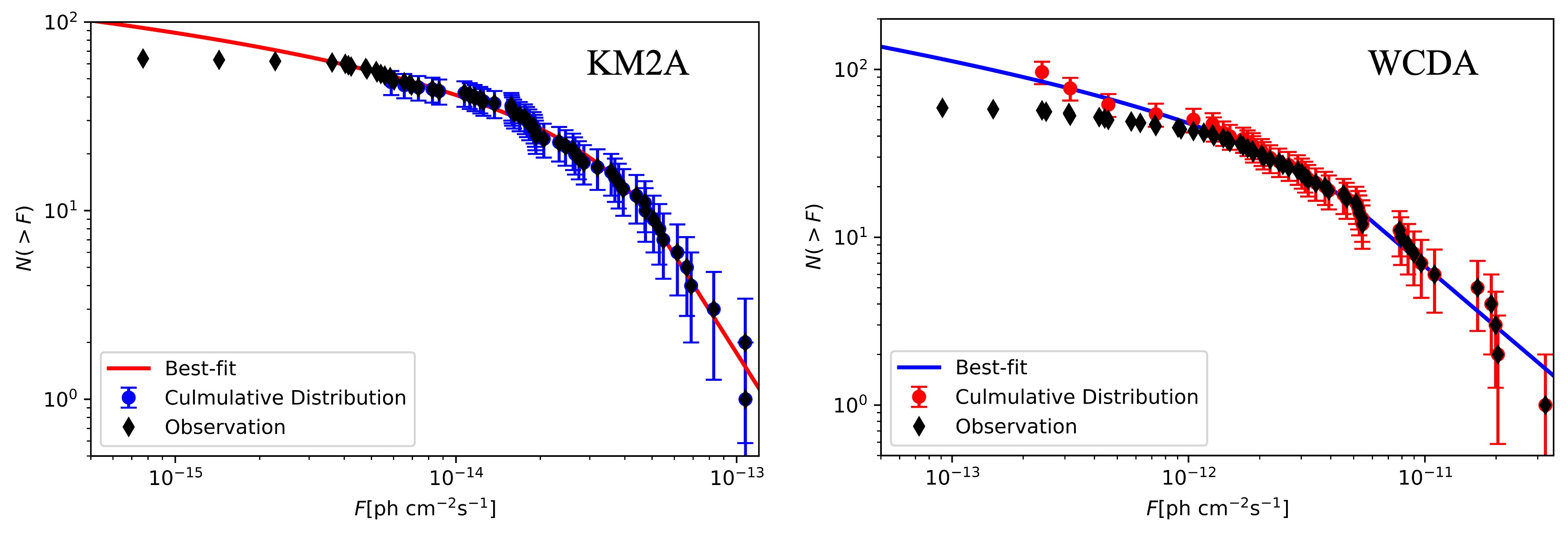}
 \caption{The cumulative distribution of integrated flux. Left panel: KM2A; right panel: WCDA. The data points are derived using the Lynden-Bell $C^{-}$ method \citep{LedynBell_1971}, and the solid curves represent the best-fit lines. We assume a broken power-law for the differential distribution of integrated flux, and the expected number of sources with integrated flux $>F$ is obtained by integrating the differential distribution over flux values greater than $F$. The black diamond data points represent observations directly obtained from LHAASO's catalog, specifically from sources with Galactic latitudes $|b|<5^{\circ}$.}
\label{fig:cdf_flux}
\end{figure*}

Figure \ref{fig:cdf_flux} illustrates the cumulative distribution of the source integrated flux for KM2A and WCDA samples
along with the best-fit curve. Table 1 summarizes the results of the fitting process. The error of $N(>F)$ are computed using \( \sigma_{N(>F)} = \frac{N(>F)}{N_{\text{obs}}(>F)} \times \sigma_{N_{\text{obs}}(>F)} \), where  \( N_{\text{obs}}(>F) \) represents the counts from data points with integrated flux above \( F \) as shown in Figure \ref{fig:gammalgF}, and \( \sigma_{N_{\text{obs}}(>F)} \) denotes the error of the observed counts, following Poisson fluctuations (i.e., \( \sigma_{N_{\text{obs}}(>F)} = \sqrt{N_{\text{obs}}(>F)} \)), consistent with the approach detailed in \citep{Singal_2012}.

\subsection{Photon spectral index distribution} \label{subsec:GAMMA}
\begin{figure*}[htb!]
\centering
\includegraphics[width=0.96\textwidth]{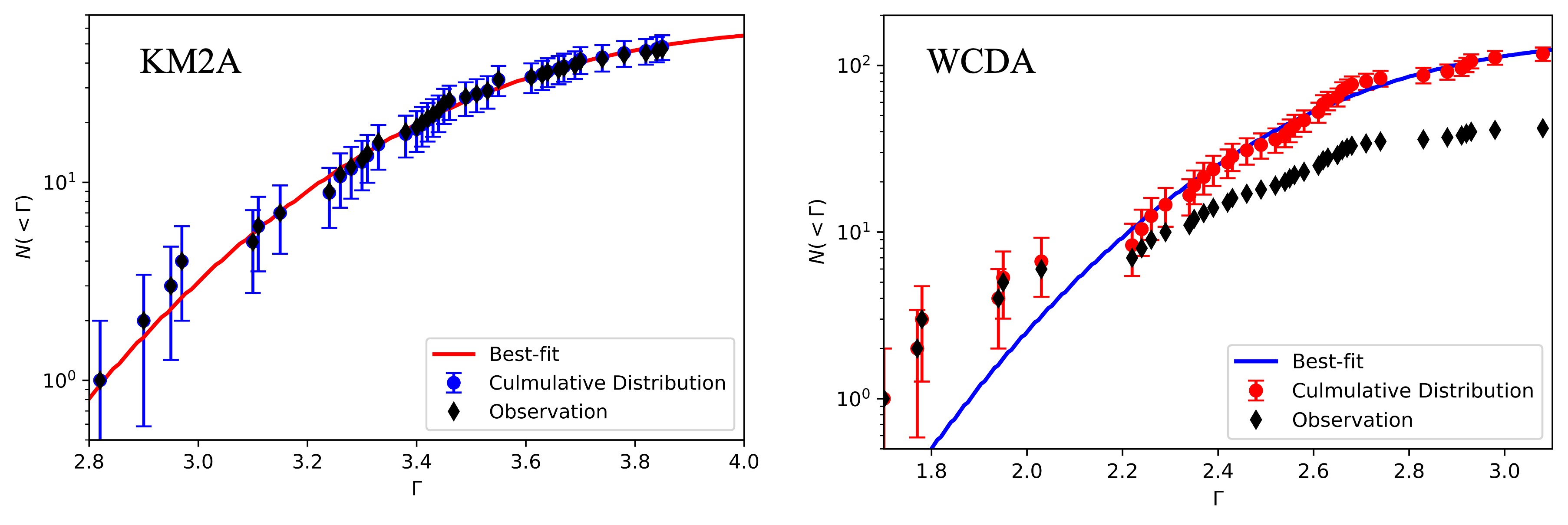}
\caption{The cumulative distribution of photon index. Left panel: KM2A; right panel: WCDA. The data points are calculated from Lyden-Bell $C^{-}$ method\citep{LedynBell_1971}, and the solid curves are the best fits. A Gaussian form is used for the description of differential distribution of photon index, where the expected number of sorces with photon index $<\Gamma$ is the integral of the differential form over photon index with $<\Gamma$. The black diamond data points represent observations that are directly obtained from our selected samples of sources that meet the detection thresholds.}
\label{fig:cdf_gamma}
\end{figure*}

The procedure described above can be applied similarly to derive the distribution of photon spectral indices. For each source above the detection threshold, a data set \( J_k' \) is defined as follows:
\begin{equation}\label{eq:index_data}
    J_{k}' = \left\{ j \mid F_{j} > F_{k}^{\text{lim}}, \Gamma_{j} \leq \Gamma_{k} \right\},
\end{equation}
where \( F_{k}^{\text{lim}} \) represents the minimum integrated flux for a source with a photon index \( \Gamma_{k} \). The dataset \( J_k' \) for the \( k \)-th source in KM2A is depicted in the left panel of Figure \ref{fig:gammalgF} by magenta square, and for WCDA, it is shown in the right panel. The number of sources in each square is denoted as \( m_k \). Following the Lyden-Bell \( C^{-} \) method, the cumulative distribution of photon indices is given by:
\begin{equation}\label{eq:psiGamma}
    \psi(\Gamma_{k}) = \prod_{j} \left( 1 + \frac{1}{m_{j}-1} \right),
\end{equation}
where the product is over all sources with \( \Gamma_{j} < \Gamma_{k} \). This formula is used to calculate the cumulative distribution of photon index, and the relationship between the cumulative distribution and the differential distribution $\frac{dN}{d\Gamma}$ is expressed as:
\begin{equation}\label{eq:intGamma}
    N(<\Gamma) \equiv \psi(\Gamma) = \int_{0}^{\Gamma} \frac{dN}{d\Gamma'} \, d\Gamma'.
\end{equation}
A Gaussian function is employed to describe the differential distribution:
\begin{equation}\label{gamma_diff}
    \frac{dN}{d\Gamma} \propto e^{-\frac{(\Gamma - \mu)^{2}}{2\sigma^{2}}},
\end{equation}
where \( \mu \) denotes the mean value and \( \sigma \) the width. The Gaussian function is normalized to unity over the  $\Gamma$ interval from 1 to 5, as it is used to compute the flux from unresolved sources in the next section.The cumulative distributions of photon indices derived from Equation \ref{eq:psiGamma} are presented in Figure \ref{fig:cdf_gamma}, and the corresponding best fits from Equation \ref{gamma_diff} are depicted as solid curves in the respective panels. The fitting results are summarized in Table \ref{tab:table_fit} as well. Figure \ref{fig:cdf_gamma} illustrates five points deviating from the best fit for WCDA, indicating a potential mixture of source populations. Specifically, sources with very hard spectral indices \( \Gamma \lesssim 2 \) at energies \( E < 25 \) TeV may have different origins compared to those with softer indices. 

\section{Estimation of fluxes of unresolved sources} 
In this section, we determine the detection efficiency of KM2A and WCDA based on the intrinsic distribution of integrated flux of the sources. Given that source masking was applied during the measurement of diffuse emission by LHAASO-KM2A, we adopt the same mask to ensure consistency with the region of interest (ROI) defined in \citep{PhysRevLett.131.151001}. Subsequently, we calculate the fraction of sources to determine the source flux.

\subsection{Detection efficiency}
Upon deriving the differential distribution \( \frac{dN}{dF} \) in Section \ref{sec:distri}, we estimate the detection efficiency by comparing it with observed sources:
\[
\lambda(F) = \frac{\left. \frac{dN}{dF} \right|_{\text{observed}}}{\left. \frac{dN}{dF} \right|_{\text{best-fit}}}
\]
Here, \( \lambda(F) \) represents the detection efficiency as a function of the integrated flux \( F \), \( \left. \frac{dN}{dF} \right|_{\text{observed}} \) denotes the number of observed sources in flux bins, and \( \left. \frac{dN}{dF} \right|_{\text{best-fit}} \) is the number of sources in flux bins from the best-fitted broken power-law function. These bins have the same lower and upper edges as those for \( \left. \frac{dN}{dF} \right|_{\text{observed}} \).

\begin{figure*}[!htb]
\centering
\includegraphics[width=0.96\textwidth]{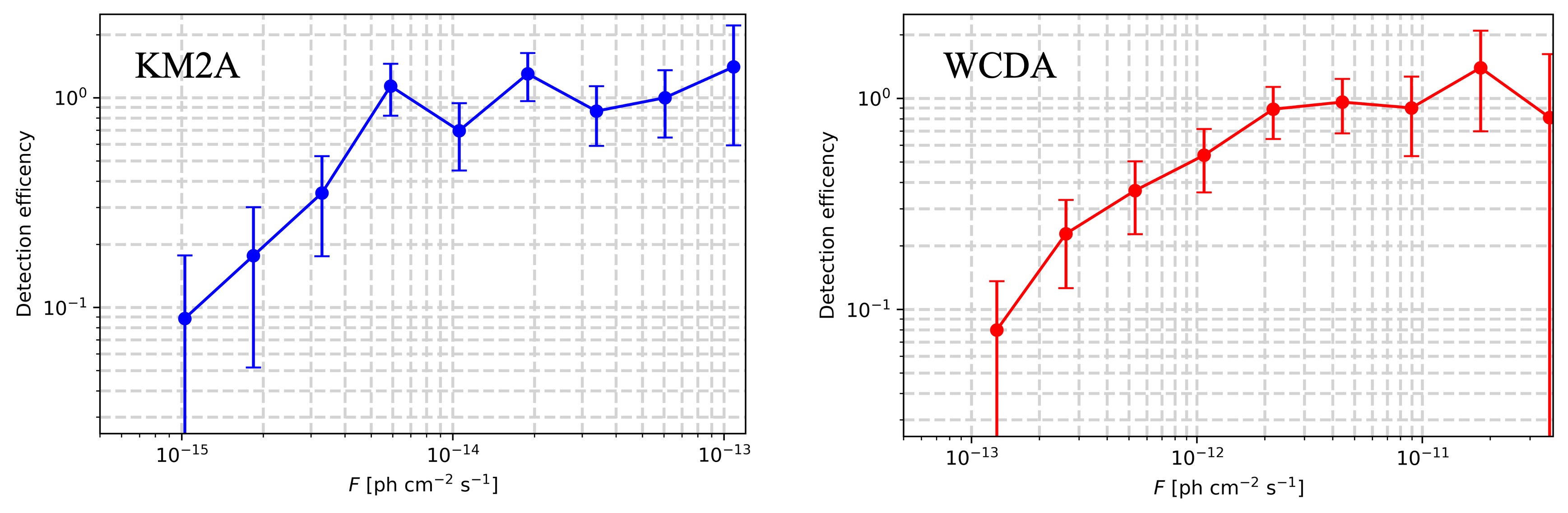}
\caption{Detection efficiencies of KM2A and WCDA, derived from the comparison between the number of observed sources and the number predicted by the source count distribution. Left panel: KM2A; right panel: WCDA.}
\label{fig:efficency}
\end{figure*}

The estimated detection efficiencies for KM2A and WCDA are illustrated in Figure \ref{fig:efficency}. Notably, for KM2A, the efficiency reaches approximately \( 100\% \) for \( F \gtrsim 6 \times 10^{-15} \, \text{ph} \, \text{cm}^{-2} \, \text{s}^{-1} \), whereas for WCDA, this threshold occurs at \( F \gtrsim 3 \times 10^{-12} \, \text{ph} \, \text{cm}^{-2} \, \text{s}^{-1} \).

\vfill

\vfill

\begin{table*}[htb!]
\centering
\caption{Fitting results of the intrinsic distributions of integrated flux and photon spectral index.}
\begin{tabular}{cccccccc}
\toprule
Sample  &  $F_0$ \footnote[1]{$F_{0}$ and $F_{b}$ is in units of $\rm ph \ cm^{-2} s^{-1}$.\label{note1}} & $\log_{10}A$ \footnote[2]{$A$ is in units of $\rm ph^{-1} \ cm^{2} s$.}  & $\beta_{1}$ & $\beta_{2}$ & $\log_{10} F_{b}$ \footref{note1} & $\mu$ & $\sigma$\\
\cline{1-8}
\midrule
KM2A & $10^{-14}$ & $15.302\pm0.077$ & $1.008\pm0.208$ & $3.348\pm0.680$ & $-13.290\pm0.101$ & $3.562\pm0.091$ & $0.343\pm0.054$\\

WCDA & $10^{-12}$ & $13.360\pm0.053$ & $1.166\pm0.152$ & $2.199\pm0.173$ & $-11.279\pm0.139$ & $2.712\pm0.064$ & $0.338\pm0.041$\\
\cline{1-8}
\bottomrule
\end{tabular}

\label{tab:table_fit}
\end{table*}

\subsection{Skymap Masking}

Given that source masking was applied during the measurement of diffuse emission by LHAASO-KM2A, it is necessary to know the spatial distribution of sources along the Galactic plane in order to calculate the flux of unresolved sources in the ROI region, based on the calculated detection efficiencies.

Assuming that the spatial distribution of sources along the Galactic plane is related to the distribution of supernova remnants or pulsars, we use the function 
\begin{equation}\label{eq:dge_dis}
f(r,z) = \left( \frac{r}{r_{\odot}} \right)^{1.25} \exp\left[-\frac{3.56(r - r_{\odot})}{r_{\odot}}\right] \exp\left( - \frac{|z|}{z_{s}}\right),
\end{equation}
to model this distribution, where $r$ is the radial distance from the Galactic center, and $z$ is the height above the Galactic plane in a cylindrical coordinate system, and $f(r,z)$ represents the source number density, expressed in units of $\rm kpc^{-3}$. $r_{\odot} = 8.5 \ \text{kpc}$ represents the radial distance of the solar system from the Galactic center, and $z_{s} = 0.2 \ \text{kpc}$ is the height of the solar system above the Galactic plane \citep{Trotta_2011}.

We calculate the expected photon flux in the line of sight from the sources of a given region using  Equation \ref{eq:dge_dis} by
\begin{equation}
N =\int _{l_{min}} ^{l_{max}} dl \int _{b_{min}} ^{b_{max}} cosb \ db \int dD \frac{f(D,l,b)D^{2}}{4\pi D^{2}}   
\end{equation}
where $f(D,l,b)$ represents the distribution in Galactic coordinates, and $D$ is the distance between a source and the Earth. We assume that the properties of each source are the same. We then use the same ROI as LHAASO-KM2A \citep{PhysRevLett.131.151001}. The fraction of photon in the ROI, either in the inner or outer Galaxy, relative to the total number of photon from sources in the Galactic plane can be estimated as

\begin{equation}\label{eq:f_i}
f_{i,\text{ROI}} = \frac{N_{i, \text{ROI}}}{N_{\text{total}}} ,  
\end{equation}
where \(i\) represents either the inner or outer Galaxy.  \(N_{\text{total}}\) is the expected total number of photon from sources on the Galactic plane (\(|b|<5^{\circ}\)) within LHAASO's field of view (right ascension \(0^{\circ}\) to \(360^{\circ}\), declination \(-20.6^{\circ}\) to \(79.4^{\circ}\)). \(N_{i, \text{ROI}}\) is the number of photon from sources in the ROI of the inner Galaxy (\(15^{\circ} < l < 125^{\circ}, |b| < 5^{\circ}\)) or outer Galaxy (\(125^{\circ} < l < 235^{\circ}, |b| < 5^{\circ}\)) used in \citep{PhysRevLett.131.151001}. Consequently, for the inner Galaxy, \(f_{\text{inner,ROI}} \approx 0.31\), and for the outer Galaxy, \(f_{\text{outer,ROI}} \approx 0.22\). The solid angles of the ROI are \(\Omega_{\text{inner,ROI}} \approx 0.206\) sr and \(\Omega_{\text{outer,ROI}} \approx 0.268\) sr, respectively.

\subsection{Flux from unresolved sources}
The flux from unresolved sources can be estimated after obtaining the intrinsic distributions of integrated flux and photon index, as well as the detection efficiency, using the following equation:

\begin{equation}\label{eq:unresolved_sed}
    S_{i}(E) = \frac{f_{i,\text{ROI}}}{\Omega_{i,\text{ROI}}} \int \phi_{0} (\frac{E}{E_{0}})^{-\Gamma} \ \frac{dN}{dF d\Gamma} (1 - \lambda(F)) dF d\Gamma
\end{equation}

Here, $i=\{\text{inner, outer}\}$, and $\phi_{0}$ is a function of $F$ and $\Gamma$ that can be calculated from Equation \ref{eq:intF}. The pivot energy $E_{0}$ is 3 TeV for WCDA and 50 TeV for KM2A, respectively. $\frac{dN}{dF d\Gamma}$ represents the number of sources in the intervals $F \sim F + dF$ and $\Gamma \sim \Gamma + d\Gamma$. The term $1 - \lambda(F)$ denotes the non-detection efficiency for unresolved sources. $f_{\rm i,ROI}$ is the photon fraction from sources of the ROI of the inner or outer Galaxy, and $\Omega_{i,\text{ROI}}$ is the solid angle of the corresponding inner or outer Galaxy region, as mentioned in the previous subsection. It should be noted that the flux in the range of $1 \sim 25$ TeV is estimated using the WCDA sample, and in the range of $25 \sim 1000$ TeV using the KM2A sample. For KM2A, the integration interval for flux is from $10^{-18} \ \text{ph} \ \text{cm}^{-2} \ \text{s}^{-1}$ to $1.5 \times 10^{-13} \ \text{ph} \ \text{cm}^{-2} \ \text{s}^{-1}$, and for the photon index from 1 to 5. For WCDA, the integration interval for flux is from $10^{-16} \ \text{ph} \ \text{cm}^{-2} \ \text{s}^{-1}$ to $5 \times 10^{-11} \ \text{ph} \ \text{cm}^{-2} \ \text{s}^{-1}$, with the same photon index range as KM2A. We have verified that varying the integration limits for the flux and photon index has a negligible effect.

\begin{figure*}[ht!]
\centering
\includegraphics[width=0.98\textwidth]{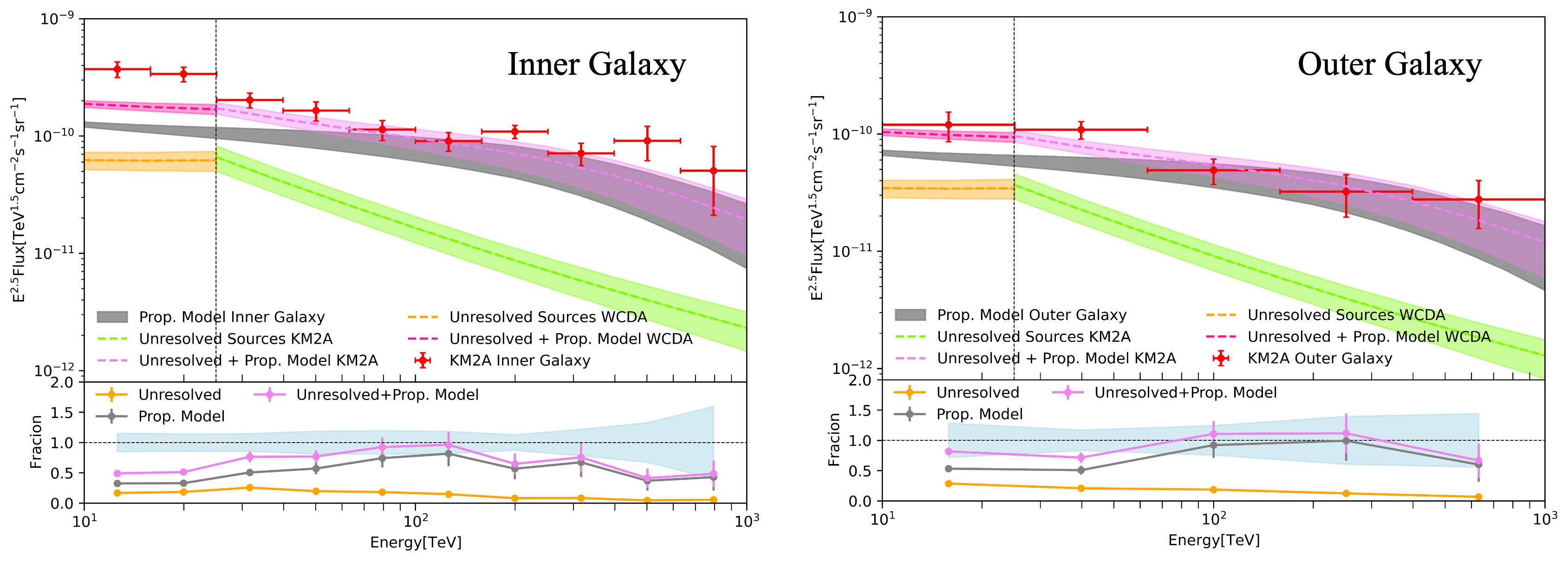}
\caption{Spectral energy distribution (SED) of unresolved sources. Left panel: inner Galaxy; right panel: outer Galaxy. Upper panel: The orange and green bands represent the flux of unresolved sources in the WCDA and KM2A energy ranges, respectively. The gray band indicates the diffuse $\gamma$-ray emission predicted by the CR propagation model \citep{Zhang_2023}. The violet and pink bands show the combined flux of unresolved sources and the propagation model in the two energy ranges. The red data points indicate the measurements \citep{PhysRevLett.131.151001}. Lower panel: Fractions relative to the measurements for different components.}
\label{fig:seds}
\end{figure*}

\section{Results}
The SEDs of unresolved sources for the inner Galaxy and outer Galaxy are presented in Figure \ref{fig:seds}. The orange and green bands represent data obtained from WCDA and KM2A, respectively. We observe a change in the spectral index between these two energy ranges, indicating that the sources observed by WCDA are harder than those observed by KM2A. Interestingly, from Equation \ref{eq:unresolved_sed}, the flux ratio of unresolved sources between the inner and outer regions is given by
$\frac{S_{\text{inner}}(E)}{S_{\text{outer}}(E)} = \frac{f_{\text{inner, ROI}} / \Omega_{\text{inner, ROI}}}{f_{\text{outer, ROI}} / \Omega_{\text{outer, ROI}}} \sim 1.8$.

The flux predicted by the CR propagation model is depicted as gray bands \citep{Zhang_2023}, with uncertainties mainly arising from the local spectra of protons and helium. In this model, CRs are assumed to propagate diffusively in a cylindrically symmetric halo, with possible convective transport away from the Galactic plane. The main propagation and source injection parameters are derived from fitting to locally measured primary and secondary spectra. The $\gamma$-ray emission is then calculated based on interactions of the equilibrium distribution of CRs with the interstellar medium and/or radiation field, including inelastic $pp$ collisions of hadronic CRs, inverse Compton scattering, and bremsstrahlung emission of electrons and positrons. The total contribution of the CR propagation model prediction and the unresolved sources is displayed as violet and pink bands. 
The fractions of the unresolved sources, the CR model prediction, and their combined 
contribution to the observed diffuse emission are shown in the bottom subplots.

For the outer Galaxy (the right panel of Figure \ref{fig:seds}), the solid black curve roughly matches within the uncertainty, indicating that the flux from unresolved sources could account for the gap between the model's prediction and the measured flux. The contribution of unresolved sources to the measured signal decreases from approximately 28\% to 7\% as energy increases. However, for the inner Galaxy, at lower energies ($E \lesssim 30$ TeV), the unresolved sources and the model's prediction can not fully account for the observed flux, leaving about 50\% of the measured flux unexplained. The contribution of unresolved sources decreases from approximately 17\% to 5\% with increasing energy.

\section{Conclusion and Discussion}
In this paper, we used the published LHAASO catalog to estimate the contribution from the unresolved sources. We first estimate the intrinsic distributions of integrated flux and photon spectral index using a non-parametric method based on the source catalog. We then derived the detection efficiencies for KM2A and WCDA, assumed a source spatial distribution, and applied the same KM2A mask to our source map. As a result, the flux contribution of unresolved sources explains the SED difference between the conventional model's predictions in the outer Galaxy's measurements. In contrast, the contribution from undetected sources is insufficient for the inner Galaxy to reconcile the difference.

It is important to note that this does not imply that introducing an undetected component is unfeasible. It remains plausible if we carefully model the source population. Pulsar halos, considered the major component of the catalog \citep{Cao_2024}, have been proposed as an explanation \citep{Yan_2024}. LHAASO has reported the discovery of an ultra-high energy gamma-ray bubble, suggesting that cosmic-ray sources in our Galaxy might have such bubbles or halos \citep{LHAASOCOLLABORATION2024449}. This indicates that such sources could contribute to the diffuse gamma-ray emission (DGE) measured by LHAASO-KM2A if they are faint or significantly extended. 
A recent study suggests that photon leakage from pulsar wind nebulae (PWNe) or halos effectively narrows the gap between observations and model predictions. This component exceeds the expected contribution from interactions between cosmic rays and the interstellar medium below approximately 25 TeV in the inner Galaxy region \citep{chen2024newperspectivediffusegammaray}. From this perspective, there seems to be no need for additional contributions from unresolved sources.

We estimated the intrinsic distribution of sources, but the incompleteness of the first LHAASO catalog introduces systematic uncertainty. Faint and very extended sources were not included in the catalog, as all sources reported by LHAASO have an extension smaller than 2 degrees \citep{Cao_2024}. It is challenging to account for a source's extension in the Lynden-Bell method for hidden correlations with integrated flux or photon index, necessitating a proper model. Moreover, the limitations of the LHAASO source survey may affect our analysis, as LHAASO only covers part of the Galactic disk. Therefore, we need to determine whether there is variation in the source population along the entire Galactic plane; the distribution should be understood as the average distribution within LHAASO's field of view. Additionally, different source populations are mixed in the same data sample, preventing us from estimating detection efficiency if we divide the sample into sub-samples, as the selection effects help us calculate the efficiency below complete detection.

Nevertheless, using the non-parametric method, estimating the contribution of unresolved sources to the diffuse gamma-ray emission measured by LHAASO-KM2A provides a new perspective on understanding the source population in the Galactic plane in the TeV to PeV domain. As LHAASO conducts more in-depth studies on cosmic-ray sources and with the construction and operation of the next generation of experiments, supplemented by multi-messenger observations, we will gain more information about these sources. If unresolved sources cannot fill the gap, the extra components will indicate hidden physics related to the origin of cosmic rays.


\begin{acknowledgments}
This work is supported by the National Natural Science Foundation of China (Nos. 12220101003,
12273114), the Project for Young Scientists in Basic Research of Chinese Academy of
Sciences (No. YSBR-061), the Natural Science Foundation for General Program of Jiangsu Province of China (No. BK20242114), and the Program for Innovative Talents and Entrepreneur in Jiangsu.

\end{acknowledgments}


\bibliography{ms}{}
\bibliographystyle{aasjournal}



\end{document}